%% file: SOCRev_Main.tex
\documentclass[twoside,english,aps,prl,superscriptaddress,twocolumn,notitlepage,10pt]{revtex4-1}
\usepackage{mathpazo}

\usepackage[LGR,T1]{fontenc}
\usepackage[latin9]{inputenc}
\usepackage[a4paper]{geometry}
\usepackage{xcolor}
\usepackage{pdfcolmk}
\usepackage{babel}
\usepackage{verbatim}
\usepackage{textcomp}
\usepackage{amsbsy}
\usepackage{amstext}
\usepackage{amssymb}
\usepackage{graphicx}
\usepackage{esint}
\PassOptionsToPackage{normalem}{ulem}
\usepackage{ulem}
\usepackage[unicode=true,pdfusetitle,
 bookmarks=false,
 breaklinks=true,pdfborder={0 0 0},pdfborderstyle={},backref=false,colorlinks=true]
 {hyperref}
\hypersetup{
 colorlinks=true,citecolor=blue,linkcolor=blue,urlcolor=blue}

\makeatletter

\DeclareRobustCommand{\greektext}{%
  \fontencoding{LGR}\selectfont\def\encodingdefault{LGR}}
\DeclareRobustCommand{\textgreek}[1]{\leavevmode{\greektext #1}}
\ProvideTextCommand{\~}{LGR}[1]{\char126#1}

\newcommand{\lyxmathsym}[1]{\ifmmode\begingroup\def\b@ld{bold}
  \text{\ifx\math@version\b@ld\bfseries\fi#1}\endgroup\else#1\fi}

\providecolor{lyxadded}{rgb}{0,0,1}
\providecolor{lyxdeleted}{rgb}{1,0,0}

\usepackage{graphicx}
\usepackage{titlesec}
\usepackage[T1]{fontenc}

\setcitestyle{super}
\titleformat{\section}[hang]{\large\bfseries\centering}{\thesection.}{1em}{}
\titlespacing\section{0pt}{4pt plus 2pt minus 2pt}{0pt plus 2pt minus 2pt}

\makeatletter
\renewcommand\frontmatter@abstractwidth{\dimexpr\textwidth-3.5cm\relax}
\makeatother
\addto\captionsenglish{}
\AtBeginDocument{
\renewcommand{\ref}[1]{\autoref{#1}}

}

\makeatother

\begin{document}

\title{Emergent Phenomena Induced by Spin-Orbit Coupling at Surfaces and
Interfaces\medskip{}
}

\author{Anjan Soumyanarayanan}

\affiliation{Division of Physics and Applied Physics, School of Physical and Mathematical
Sciences, Nanyang Technological University, 637371 Singapore}

\affiliation{Data Storage Institute, 2 Fusionopolis Way, 138634 Singapore}

\author{Nicolas Reyren}

\affiliation{Unité Mixte de Physique, CNRS, Thales, Univ. Paris-Sud, Université
Paris-Saclay, Palaiseau 91767, France}

\author{Albert Fert}
\email{albert.fert@thalesgroup.com}

\affiliation{Unité Mixte de Physique, CNRS, Thales, Univ. Paris-Sud, Université
Paris-Saclay, Palaiseau 91767, France}

\author{Christos Panagopoulos}
\email{christos@ntu.edu.sg}

\affiliation{Division of Physics and Applied Physics, School of Physical and Mathematical
Sciences, Nanyang Technological University, 637371 Singapore}
\begin{abstract}
\textbf{Spin-orbit coupling (SOC) describes the relativistic interaction
between the spin and momentum degrees of freedom of electrons, and
is central to the rich phenomena observed in condensed matter systems.
In recent years, new phases of matter have emerged from the interplay
between SOC and low dimensionality, such as chiral spin textures and
spin-polarized surface and interface states. These low-dimensional
SOC-based realizations are typically robust and can be exploited at
room temperature (RT). Here we discuss SOC as a means of producing
such fundamentally new physical phenomena in thin films and heterostructures.
We put into context the technological promise of these material classes
for developing spin-based device applications at RT. }
\end{abstract}
\maketitle

\input{SOCRev_A-Intro.tex}

\input{SkTuning_B-SPStates.tex}

\input{SOCRev_C-Magnetism.tex}

\input{SOCRev_D-Impact.tex}

\noindent \bibliographystyle{naturemag}
\bibliography{SOC-Rev}

\noindent \begin{center}
\rule[0.5ex]{0.6\columnwidth}{1pt}
\par\end{center}

\noindent \textbf{\emph{\small{}Acknowledgments.}}{\small{} We are
grateful to A.K.C. Tan and S.M. Rezende for their help preparing illustrations.
We acknowledge support from the Singapore Ministry of Education (MOE),
an Academic Research Fund Tier 2 (Reference No. MOE2014-T2-1-050),
the National Research Foundation (NRF) of Singapore, a NRF Investigatorship
(Reference No. NRF-NRFI2015-04) and the A{*}STAR Pharos Fund (1527400026),
Singapore; and the Centre National de la Recherche Scientifique (CNRS),
France, for funding this work. }{\small \par}

\noindent \textbf{\emph{\small{}Author Contributions.}}\textbf{\small{}
}{\small{}All authors contributed equally to this work. }{\small \par}

\noindent \clearpage{}
\end{document}

%% file: SOCRev_A-Intro.tex
\section*{Introduction}

\noindent 
\begin{figure*}
\begin{centering}
\includegraphics[width=6.9in]{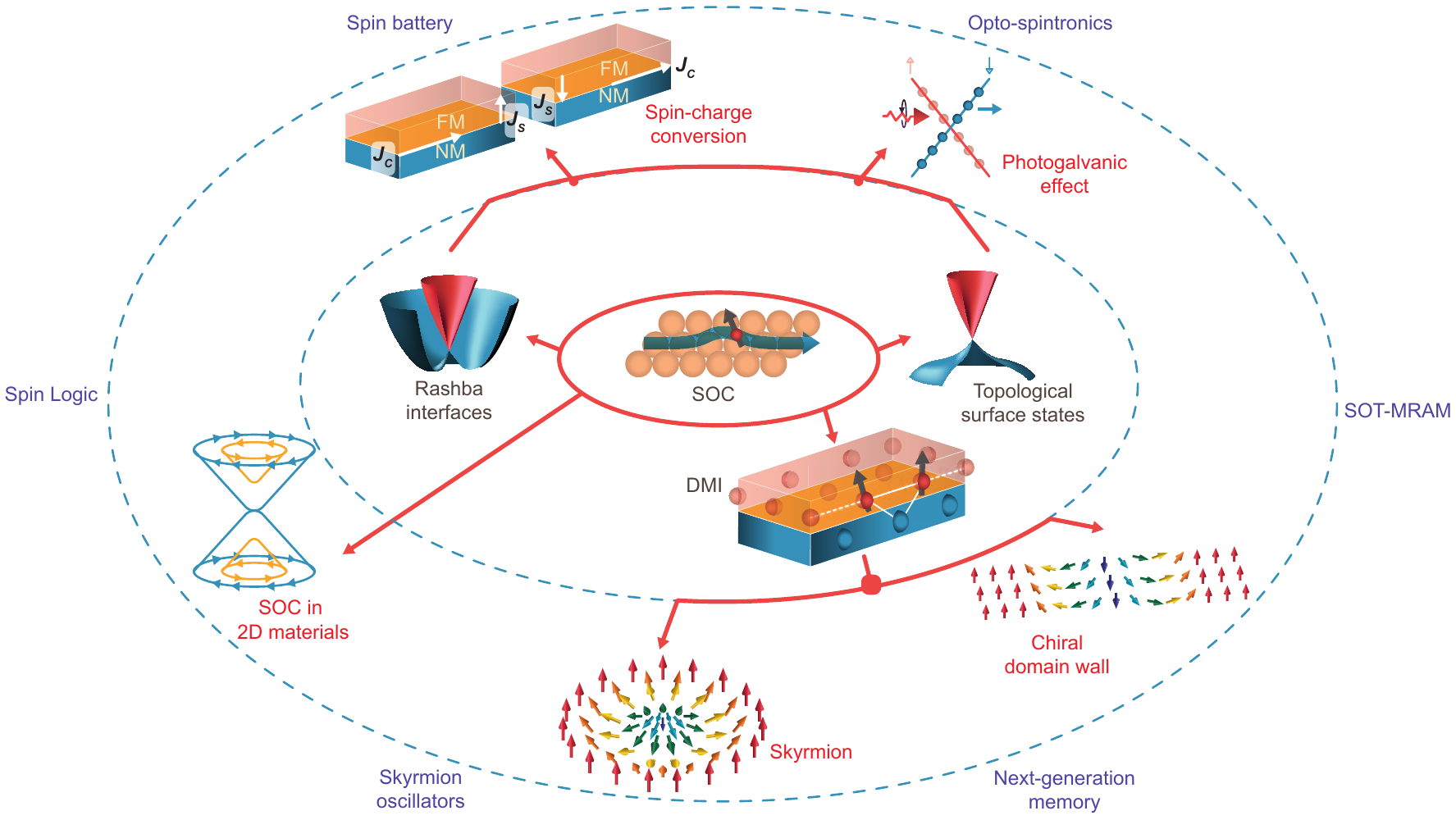}
\par\end{centering}
\caption{\textbf{Emergent Phenomena from Spin\textendash Orbit Coupling (SOC)
at Surfaces and Interfaces. }A schematic illustration of the connection
between the presence of strong SOC at material surfaces and interfaces
(inner ellipse) and the resulting emergence of new interactions and
electronic states (middle ellipse), such as Dzyaloshinskii\textendash Moriya
interaction (DMI; see \ref{fig:Sk-Form}\textcolor{blue}{a, e} for
details), Rashba interfaces (\textcolor{blue}{\ref{fig:SP-Bands}b,
d}) and topological surface states (\textcolor{blue}{\ref{fig:SP-Bands}a,
c}). These emergent phenomena can in turn be used to generate new
2D spintronics effects (outer ellipse), such as spin\textendash charge
conversion (\textcolor{blue}{\ref{fig:SP-Bands}e, f} and \textcolor{blue}{\ref{fig:SC-Conv}}),
the photogalvanic effect, enhanced  SOC in 2D materials, such as graphene
(\textcolor{blue}{\ref{fig:SC-Conv}d, e}), magnetic skyrmions (\ref{fig:Sk-Form}\textcolor{blue}{b})
and chiral domain walls (\ref{fig:Sk-Form}\textcolor{blue}{c}), which
have direct device applications (periphery). FM, ferromagnet; NM,
non-magnetic material.\label{fig:SOC-Schematic}}
\end{figure*}
The electric field experienced by a travelling electron translates,
in its rest frame, to a magnetic field proportional to its velocity
\textendash{} a relativistic effect which is notable in crystalline
lattices with heavy atoms. The Zeeman interaction between the electron
spin and this effective magnetic field is equivalent to the coupling
of the electronic spin and momentum degrees of freedom, known as SOC.
SOC can split degenerate bands with finite angular momentum ($p$,
$d$ and $f$), modifying the electronic band structure. Importantly,
SOC effects are greatly enhanced in reduced dimensions (\ref{fig:SOC-Schematic},
left and right). First, inversion symmetry is broken at the surface
or interface, and the resultant electric field couples to the spin
of itinerant electrons. This phenomenon, known as Rashba SOC\citep{Rashba1960},
produces spin-split dispersion even at the\- surfaces of conventional
metals (such as Au and Bi)\citep{Manchon2015}. Recently discovered
topological insulators (TIs), have spin-polarized surface states with
additional topological properties. In both these cases, strong two-dimensional
(2D) SOC locks the electron spin and momentum.

Spin\textendash momentum locking in 2D geometries has direct consequences
for the interplay between the charge and spin transport (\ref{fig:SOC-Schematic},
top left). An in-plane charge current induces a transverse spin accumulation
(uniform non-zero spin density). This spin accumulation can be used
to eject a spin current into an adjacent layer (Edelstein effect\citep{Edelstein1990}).
Conversely, the injection of a spin current induces the associated
spin polarization and charge current in the 2D states. Other types
of conversion between charge and spin currents can also be obtained
by SOC effects in three-dimensional (3D) conductors, namely the spin
Hall effect of heavy metals\citep{Hoffmann2013}; however, the observed
effects in two dimensions are considerably enhanced. Such spin\textendash charge
conversion phenomena have direct applications for spintronics technologies,
which are based on the creation and detection of spin currents\citep{Valenzuela2012}.
Given that the 3D spin Hall effect is already used in spintronics
devices\citep{Cubukcu2014}, the observed effects in two dimensions
offer much promise for device applications (\ref{fig:SOC-Schematic},
top).

The interplay between SOC and magnetism is of increasing importance.
In conventional magnetic materials, ferromagnetic order, which results
from exchange interaction, aligns neighbouring spins. A well-known
consequence of SOC is magneto-crystalline anisotropy \textendash{}
the preferential alignment of electron moments along certain crystallographic
directions (`easy axes\textquoteright ), via the coupling of electron
motion to the crystalline lattice field. In systems that lack inversion
symmetry, SOC induces a chiral Dzyaloshinskii\textendash Moriya interaction
(DMI)\citep{Dzyaloshinsky1958a,Moriya1960}, which takes the form:

\noindent 
\begin{equation}
\mathcal{H}_{{\rm DM}}=-(\boldsymbol{S}_{1}\times\boldsymbol{S}_{2})\cdot\boldsymbol{D}_{12}\label{eq:DMI}
\end{equation}

Here $\boldsymbol{S}_{1}$ and $\boldsymbol{S}_{2}$ are neighbouring
spins and $\boldsymbol{D}_{12}$ is the Dzyaloshinskii\textendash Moriya
vector. The DMI is a chiral interaction that decreases or increases
the energy of the spins depending on whether the rotation from $\boldsymbol{S}_{1}$
to $\boldsymbol{S}_{2}$ around $\boldsymbol{D}_{12}$ is clockwise
or anticlockwise. If $\boldsymbol{S}_{1}$ and $\boldsymbol{S}_{2}$
are initially parallel, then the effect of a sufficiently strong DMI
(with respect to exchange and anisotropy) is to introduce a tilt around
$\boldsymbol{D}_{12}$. DMI was initially understood as a super-exchange
interaction in magnetic insulators\citep{Dzyaloshinsky1958a,Moriya1960},
and later extended to non-centrosymmetric magnetic metals\citep{Levy1980}.
In a disordered magnetic alloy, a large SOC element could mediate
such an interaction between two nearby magnetic atoms, with the resulting
Dzyaloshinskii\textendash Moriya vector being perpendicular to the
plane formed by the three atoms. Crucially, this model was extended
to magnetic multilayers, wherein inversion symmetry is broken by the
presence of an interface\citep{Fert1990} (\ref{fig:SOC-Schematic},
bottom right). The existence of interfacial DMI was first demonstrated
by the observation of spiral-like spatial modulations of the spin
orientation with a winding periodicity related to the magnitude of
the DMI\citep{Bode2007}. DMI also enables the formation of other
chiral spin structures \textendash{} in particular, chiral domain
walls and skyrmions \textendash{} that are possibly relevant to next-generation
information storage devices (\ref{fig:SOC-Schematic}, bottom).

Recent developments in the techniques for thin-film growth and in
the capabilities of ab initio calculations have enabled the synthesis
of atomically flat surfaces and heterostructures, and the prediction
of their electronic properties. A common thread across several such
thin-film materials and heterostructures \textendash{} heavy metal
compounds and multilayers \textendash{} is that the SOC strength at
surfaces and interfaces is comparable to the other relevant energy
scales, and so plays a pivotal part. In combination with surface and
interface effects, this engenders fundamentally new spin-based phenomena
that are robust to disorder and thermal fluctuations, with much promise
for RT spin-based applications.

Here we describe these diverse low-dimensional spin-based phenomena
in the context of their SOC origin. We begin by detailing the progress
on spin-polarized states at the surfaces of TIs, Rashba interfaces
and atomically thin (2D) materials, and examine their utility towards
the generation and conversion of spin currents. Next, we describe
the developments on interfacial-DMI-induced non-collinear spin textures
\textendash{} skyrmions and chiral domain walls (DWs) \textendash{}
in magnetic films, and techniques to generate, stabilize and manipulate
them in devices. Finally, we explore the feasibility of realizing
the technological promise of these diverse SOC induced surface and
interface phenomena towards RT device applications.

%% file: SkTuning_B-SPStates.tex
\section*{Spin-Polarized Surface\protect \\
 and Interface States}

\noindent 
\begin{figure*}
\begin{centering}
\includegraphics[width=5.6in]{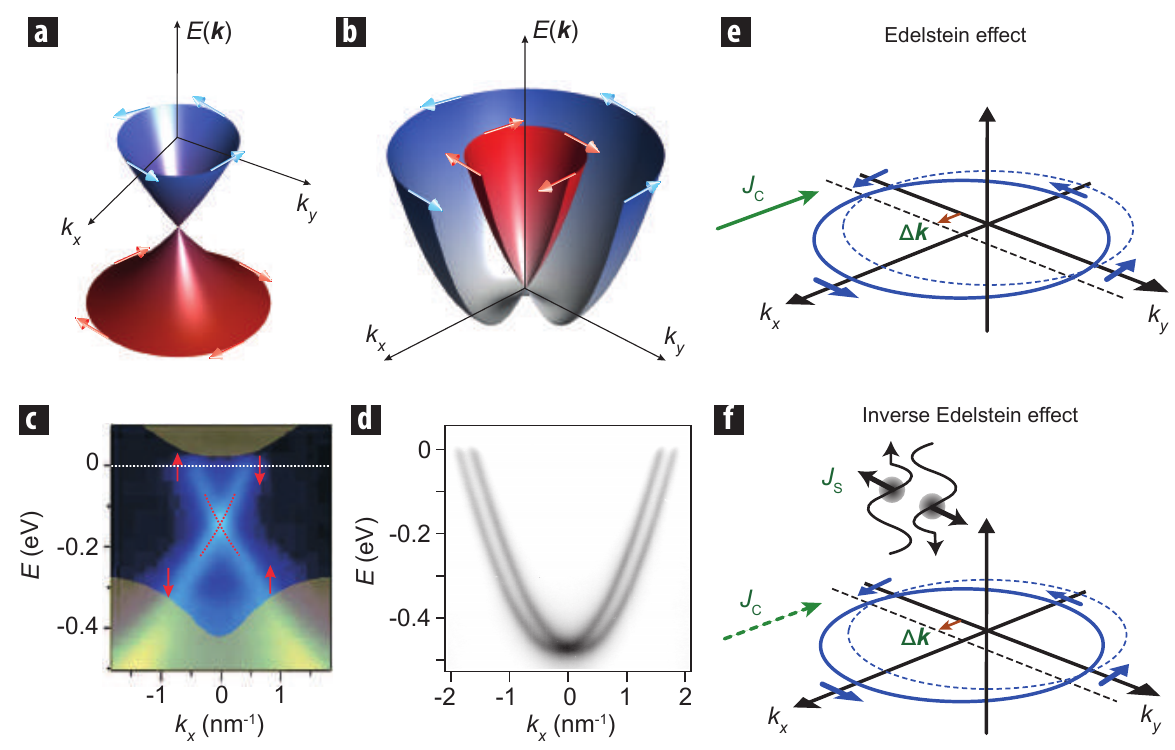}
\par\end{centering}
\caption{\textbf{Band Structure and Spin-Charge Conversion in Spin-Polarized
2D States.} \textbf{(a, b)} Schematic of the spin-polarized band structure
(electron energy $E$ as a function of in-plane momentum $k$) of
2D electron states\citep{Valenzuela2012} at the surfaces and interfaces
of TIs (a) and at Rashba systems (b). The arrows indicate electron
spin, with blue and red dispersion surfaces corresponding to opposite
spin helicities. \textbf{(c, d)} ARPES measurements of the 2D band
structure, with respect to the Fermi level ($E=0$), at the surface
of the TI Bi$_{2-x}$Ca$_{x}$Se$_{3}$\citep{Hsieh2009c} (c) and
at the Rashba surface of Au(111)\citep{Nechaev2009} (d). The red
arrows in (c) indicate the spin orientation, and the bulk bands are
schematized in brown. \textbf{(e)} $k$-space schematic of charge-to-spin
conversion in TIs via the Edelstein effect (EE). A charge current
$J_{{\rm c}}$ at the surface causes a shift $\Delta k$ of the Fermi
contour, resulting in a non-zero spin density as a result of the helical
spin orientation. This spin density can diffuse as a spin current
in the adjacent material. \textbf{(f)} Spin-to-charge conversion by
the inverse Edelstein effect (IEE) in TIs. Injecting a spin current
($J_{s}$; spin-polarized wiggles) into the surface states of the
TI overpopulates states on one side of the Fermi contour and depopulates
states on the other, generating a charge current.\label{fig:SP-Bands}}
\end{figure*}

\noindent \textsf{\textbf{Rashba States}}

\noindent The Rashba effect arises from SOC and broken inversion symmetry
at material surfaces and interfaces\citep{Rashba1960}, with the corresponding
Hamiltonian: 

\noindent 
\begin{equation}
\mathcal{H}_{{\rm R}}=v_{0\,}\hat{\boldsymbol{z}}\cdot(\boldsymbol{k}\times\boldsymbol{\sigma})\label{eq:2D-SP-States}
\end{equation}

\noindent Here $v_{0}$ is the Rashba parameter, $\boldsymbol{\sigma}$
is spin, $\boldsymbol{k}$ is momentum and $\hat{\boldsymbol{z}}$
is the unit normal to the surface or interface. The Rashba effect
results in spin-split 2D dispersion surfaces and, importantly, in
the locking of spin and momentum degrees of freedom to each other
(\ref{fig:SP-Bands}\textcolor{blue}{b}). 

Rashba SOC\textendash split states have been investigated across various
surfaces and interfaces\citep{Manchon2015,LaShell1996,Nechaev2009},
as shown for angle-resolved photoemission spectroscopy (ARPES) measurements
of the Au(111) surface (\ref{fig:SP-Bands}\textcolor{blue}{d})\citep{Nechaev2009}.
Interface alloying of heavy elements with intermediate-weight metals
can enhance the in-plane potential gradient via hybridization, leading
to more pronounced Rashba effects, as on the Bi/Ag(111) alloyed interface
($v_{0}=3$~eV·Å\citep{Ast2007a}).

\noindent \textsf{\textbf{Topological Surface States}}

\noindent In materials with heavy elements, strong SOC can split the
$p$ band by a large enough magnitude to flip the $s\lyxmathsym{\textendash}p$
band structure, inducing band inversion. Notably, 2D heterostructures
of Hg$_{x}$Cd$_{1-x}$Te exhibit, in addition to such an inversion,
an associated topological phase transition\citep{Bernevig2006a,Koenig2007},
which results in protected states at the edges of the sample. The
presence of edge states in 2D heterostructures was subsequently generalized
to 3D insulators\citep{Fu2007,Hasan2010}. The surfaces or interfaces
of such TIs must host protected states at time-reversal-invariant
$k$-space points\citep{Fu2007,Hasan2010}. These topological surface
states have a nearly linear energy\textendash momentum relationship
(\ref{fig:SP-Bands}\textcolor{blue}{a})\citep{Hasan2010}. The Dirac
Hamiltonian that describes these surface states, $\mathcal{H}_{D}=v_{0\,}\hat{\boldsymbol{z}}\cdot(\boldsymbol{k}\times\boldsymbol{\sigma})$,
has the same Rashba form (\ref{eq:2D-SP-States}) and locks the spin
and momentum degrees of freedom (\ref{fig:SP-Bands}\textcolor{blue}{a,
c})\citep{Hasan2010}. However, whereas Rashba SOC leads to spin-split
parabolic surface states in conventional metals, topological surface
states are distinguished by their helical single Dirac cone character,
which emerges from the requirement to connect the bulk valence and
conduction bands.

ARPES measurements demonstrated the topological nature of surface
states first in the indirect band gap semiconductor Bi$_{1-x}$Sb$_{x}$\citep{Hsieh2008}
and then in a larger, direct band gap (300 meV) TI Bi$_{2}$Se$_{3}$\citep{Hsieh2009c}.
The discovery of a simple Dirac cone within the band gap of bulk Bi$_{2}$Se$_{3}$
(\ref{fig:SP-Bands}\textcolor{blue}{c}), with a chemical potential
that is tunable via chemical doping\citep{Hsieh2009c} and the electric
field effect\citep{Checkelsky2011}, has since led to the discovery
of several other single-Dirac-cone TIs\citep{Hasan2010}. 

The electronic transport of TIs is governed by the helical Dirac nature
of topological surface states. First, surface-state transport arises
from a 2D Dirac cone: therefore, it can be ambipolar, controlled by
electric fields, and tuned through the Dirac point with a characteristic
minimum conductivity\citep{Checkelsky2011}. Second, spin\textendash momentum
locking prevents backscattering between states of opposite momenta
with opposite spins, as evidenced across several TIs\citep{Beidenkopf2011}.
Because backscattering dominates charge dissipation in conventional
metals, quasiparticles of TIs are expected to exhibit longer lifetimes,
enabling ballistic phenomena such as the quantum Hall effect\citep{Brune2011}.
Third, the polarization of light incident on a TI can couple to the
surface-state momentum, thereby generating spin-polarized photocurrents
with high fidelity\citep{McIver2011}. Finally, magnetic doping of
TIs breaks time-reversal symmetry, leading to a gap at the Dirac point\citep{Chen2010a}
and a reorientation of the low-energy spin texture\citep{Xu2012e}.
The mediation of magnetism by Dirac fermions in TIs enables exotic
phenomena, such as the quantum anomalous Hall effect\citep{Chang2013}.

\noindent 
\begin{figure*}
\begin{centering}
\includegraphics[width=6.9in]{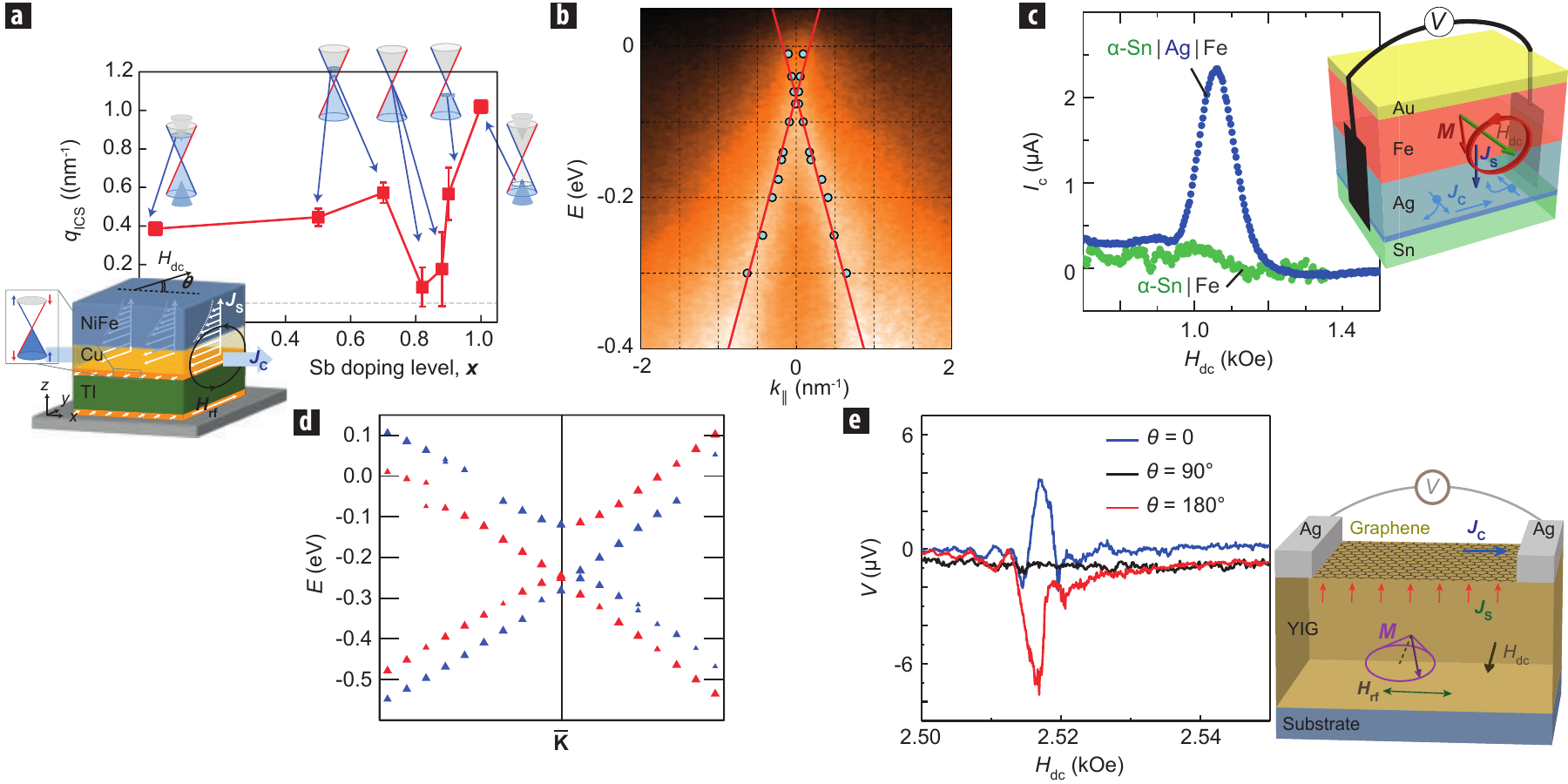}
\par\end{centering}
\caption{\textbf{Spin\textendash Charge Conversion Experiments.} \textbf{(a)}
Charge-to-spin conversion by EE in spin torque ferromagnetic resonance
(ST-FMR) experiments on the TI (Bi$_{1-x}$Sb$_{x}$)$_{2}$Te$_{3}$
at $T=10$~K\citep{Kondou}. The graph shows the spin-charge conversion
coefficient $q_{{\rm ICS}}$ for several Sb concentrations $x$, and
the corresponding Fermi levels in the Dirac cone. Inset, schematic
of the device, in which an applied longitudinal a.c. charge current
$J_{{\rm c}}$ (associated with an a.c. magnetic field $H_{{\rm rf}}$)
is converted by EE into a vertical spin current $J_{s}$. This spin
current is injected through a Cu layer into the top NiFe layer and
detected by a ST-FMR-induced d.c. voltage, in an external d.c. field
$H_{{\rm dc}}$ applied at an angle $\theta$. \textbf{(b, c)} Spin-to-charge
conversion by IEE in spin-pumping experiments on the TI \textgreek{a}-Sn\citep{Rojas-Sanchez2016}.
(b) ARPES intensity maps (for varying electron energy $E$ and in-plane
momentum $k_{\parallel}$) show that the surface-state Dirac cone
on \textgreek{a}-Sn is preserved (symbols on the red lines represent
maxima in ARPES intensity scans) even after Ag deposition (to a thickness
of 2.3~nm). c, A d.c. voltage generated by IEE is observed when \textgreek{a}-Sn
is covered by Ag (blue), but not when it is directly covered by Fe,
which destroys the Dirac cone (green). Inset, a FMR spin-pumping device
(the magnetization $M$ of the ferromagnetic layer in an external
field $B=\mu_{0}H_{{\rm dc}}$, excited by an a.c. field) injects
a vertical spin current $J_{s}$ through an Ag layer into the surface
states of the TI, which generates a charge current $I_{c}$. \textbf{(d)}
Calculated band structure of graphene after Au intercalation between
graphene and substrate, matched to ARPES measurements near the ${\rm K}$-point\citep{Marchenko2012}.
Blue and red symbols indicate opposite spin orientations derived from
fits to spin-resolved ARPES measurements. \textbf{(e)} Spin-to-charge
conversion by spin pumping from yttrium iron garnet (YIG) into graphene\citep{Mendes2015}:
experimental set-up (inset) and lateral voltage $V$ induced by IEE
for opposite applied fields.\label{fig:SC-Conv}}
\end{figure*}

\noindent \textsf{\textbf{Conversion between Spin and Charge Currents}}

\noindent The helical spin polarization of Fermi contours of Rashba
interfaces and surfaces of TIs enables the conversion between spin
and charge currents by the Edelstein and inverse Edelstein effects\citep{Edelstein1990}
(EE and IEE, \ref{fig:SP-Bands}\textcolor{blue}{e, f}). In a single-cone
TI, a charge current\textemdash that is, a shift of the Fermi contours
in the direction of the electron motion ($x$) (\ref{fig:SP-Bands}\textcolor{blue}{e})
\textendash{} induces an overpopulation of spins in the transverse
direction ($y$) as a result of spin\textendash momentum locking,
and is therefore associated with a nonzero spin accumulation. The
spin accumulation can diffuse through an interface into an adjacent
conducting material, resulting in a pure 3D spin current being injected
into this material, without a net charge flow. The spin\textendash charge
conversion yield is quantified by the inverse length $q_{{\rm ICS}}$
\textendash{} the ratio between the resultant 3D spin current density
and the applied 2D charge current density\citep{Kondou}. Furthermore,
if the spin current is injected into a magnetic material, then the
resulting spin-transfer torque can be used to switch its magnetization.
Alternatively, an a.c. charge current could be used to induce ferromagnetic
resonance. 

In the inverse conversion of a spin current into a charge current
by IEE on the interface of a TI (\ref{fig:SP-Bands}\textcolor{blue}{f}),
the injection (extraction) of spins oriented along $+y$ ($-y$) into
(from) the states of a helically spin-polarized Fermi contour populates
( depopulates) states on the $+x$ ($-x$) side of the Fermi contour.
This out-of-equilibrium distribution corresponds to a charge current.
The conversion coefficient \textendash{} the ratio between the induced
2D charge current density and applied 3D spin current density \textendash{}
is the IEE length, $\lambda_{{\rm IEE}}$\citep{Sanchez2013,Shen2014}.
For a pure helical ground state $\lambda_{{\rm IEE}}=v_{{\rm F}}\tau$\citep{Rojas-Sanchez2016},
where $v_{{\rm F}}$ is the Fermi velocity and $\tau$ is the relaxation
time of an out-of-equilibrium distribution in the interface states.
The spin\textendash charge conversion by EE and IEE at Rashba interfaces
can be described in an analogous fashion, but accounting for the partial
compensation of the two Fermi contours of opposite chirality, yielding
$\lambda_{{\rm IEE}}=v_{0}\tau/\hbar$\citep{Sanchez2013,Shen2014},
where $\hbar$ is the reduced Planck constant.

Examples of TIs in which charge-to-spin conversion effects have been
observed include (Bi$_{1-x}$Sb$_{x}$)$_{2}$Te$_{3}$\citep{Kondou}
(\ref{fig:SC-Conv}\textcolor{blue}{a}). Here, the charge-to-spin
conversion of the applied a.c. charge current generates a vertical
spin current, and a spin-transfer torque on the magnetization of the
deposited NiFe layer. When tuning the Fermi level across the Dirac
point by varying the doping $x$, the conversion coefficient $q_{{\rm ICS}}$
does not change sign because both charge and spin chiralities change
simultaneously (\ref{fig:SC-Conv}\textcolor{blue}{a}). However, $q_{{\rm ICS}}$
exhibits a sharp minimum at the Dirac point, at which a finite scattering
rate between the upper and lower cone can mix the spin polarizations.
Similar results of the generation of spin polarization and spin torques
have been obtained for other TIs\citep{Mellnik2014,Fan2014a,Li2014,Wang2016}.
Spin-to-charge conversion on TIs by IEE has been achieved both in
spin pumping\citep{Shiomi2014} and in tunnelling spin injection\citep{Wang2016}
experiments. In the case of spin pumping conversion (\ref{fig:SC-Conv}\textcolor{blue}{b,
c})\citep{Rojas-Sanchez2016}, a vertical spin current $J_{{\rm s}}$,
which is produced by ferromagnetic resonance from a ferromagnetic
layer (NiFe), is injected into the TI \textgreek{a}-Sn through an
intermediate Ag layer (\ref{fig:SC-Conv}\textcolor{blue}{c}). The
persistence of the Dirac cone on \textgreek{a}-Sn after Ag deposition
is verified by ARPES (\ref{fig:SC-Conv}\textcolor{blue}{b}), whereas
direct deposition of Fe on Sn destroys the Dirac cone. A charge current
induced by IEE is consistently found with Ag on \textgreek{a}-Sn (\ref{fig:SC-Conv}\textcolor{blue}{c},
blue), and not with Fe (\ref{fig:SC-Conv}\textcolor{blue}{c}, green).
Here, $\lambda_{{\rm IEE}}=2.1$~nm at RT \textendash{} a spin\textendash charge
conversion efficiency that is at least an order of magnitude higher
than that obtained with the inverse spin Hall effect in metals such
as Pt or W\citep{Rojas-Sanchez2016}.

Analogous spin\textendash charge conversions by EE and IEE can also
be obtained at Rashba interfaces\citep{Sanchez2013,Nomura2015}. However,
owing to the compensation between the two Fermi contours of the Rashba
interface, the conversion is generally not as efficient as for TIs.
An exception is the interface between the insulating oxides SrTiO$_{3}$
and LaAlO$_{3}$, for which IEE values as large as 6~nm have been
reported\citep{Lesne2016}.

\noindent \textsf{\textbf{Materials Directions and Applications}}

\noindent Despite tremendous excitement surrounding TIs, the development
of technologically relevant materials has been hindered by several
issues. In binary TIs (for example, Bi$_{2}$Se$_{3}$), intrinsic
defects pin the chemical potential within the bulk bands, diminishing
the contribution of the surface-state transport\citep{Hsieh2009c}.
Although chemical doping could tune the chemical potential into the
bandgap, this markedly reduces the mean free path\citep{Beidenkopf2011}.
Therefore, recent efforts have used epitaxial techniques to fabricate
ternary TIs with better control over the chemical potential and mobility\citep{Zhang2011}.
Other efforts towards functional TIs include systems with larger Rashba
SOC and stack engineering of strong SOC materials\citep{Wang2013}.
Another emerging direction lies at the crossover between Dirac and
Rashba behaviour, with the Dirac/Rashba character of quasiparticles
tuned by film thickness (for example, few-layer Bi2Se3 films\citep{Zhang2009})
or chemical potential (for example, Sb\citep{Soumyanarayanan2014}).

The partial freezing of backscattering for surface and interface states
of TIs, which results in reduced energy dissipation by electrical
currents, is a notable advantage for their use in lowpower nano-devices.
As a next step, the ambipolarity of topological surface states could
be exploited to create, for example, topological $p\lyxmathsym{\textendash}n$
junctions\citep{Wang2012i} and spin transistors. The spin Hall effect
of heavy metals is already used in spin\textendash orbit torque magnetic
random-access memory (SOT-MRAM) switching\citep{Cubukcu2014}, and
similar conversions by TIs or Rashba interfaces are expected to be
much more energy-efficient\citep{Mellnik2014}.

\noindent \textsf{\textbf{Enhanced SOC in 2D Materials}}

As with surface and interface states of TIs, the 2D electronic states
of single-layer graphene are characterized by linearly dispersing
Dirac cones at the ${\rm K}$ and ${\rm K'}$ points in reciprocal
space\citep{CastroNeto2009b}. In contrast to TIs, the SOC magnitude
for the $sp^{2}$ bonded structure of pristine graphene is small (about
10~\textgreek{m}eV)\citep{CastroNeto2009b}, and Dirac cones in graphene
are generally supposed to be spin-degenerate. However, SOC can be
greatly enhanced by proximity and hybridization with adjacent materials.
In-plane and out-of-plane deformations that mix the $sp^{2}$ and
$sp^{3}$ orbitals in strained or buckled graphene, as well as interactions
with adatoms and electric fields, can also lead to enhancements in
SOC\citep{CastroNeto2009,Gmitra2009,Marchenko2012,Rashba2009,Ferreira2014}.
The SOC effects of an ordered interface between graphene and another
material on the energy dispersion of graphene can be expressed as\citep{Rashba2009}

\noindent 
\begin{equation}
E_{\alpha\beta}=\frac{\alpha\lambda}{2}\,+\,\beta\,\left(k^{2}+\frac{\lambda^{2}}{4}\right)^{2}\label{eq:Graphene-SOC}
\end{equation}

with $\beta=+1(-1)$ for the upper (lower) dispersion surfaces, $\alpha=+1(-1)$
for the split upper (lower) dispersion surfaces, and where $\lambda$
is the SOC constant of the system and $k=\left|\boldsymbol{k}\right|$.
Therefore, SOC-induced splitting of graphene bands leads further to
spin\textendash momentum locking and chiral spin orientations; for
example, the large SOC splitting (about 100~meV) for graphene on
Au (see, for example, \ref{fig:SC-Conv}\textcolor{blue}{d}) is caused
by the strong hybridization between the Dirac-cone states and $d$
states of Au\citep{Marchenko2012}. 

Enhancement of SOC in graphene can similarly lead to 2D spin\textendash charge
conversion effects. For example, ferromagnetic-resonance spin pumping
from yttrium iron garnet (YIG) into graphene induces a definite broadening
and a lateral voltage characteristic of IEE (\ref{fig:SC-Conv}\textcolor{blue}{e})\citep{Mendes2015}
that is ascribed to the SOC induced by proximity with YIG. Here, $\lambda_{{\rm IEE}}=10^{-3}$~nm,
which is much smaller than with TIs or Rashba interfaces, indicating
a moderate SOC splitting. On the other hand, intercalation of, for
example, an Au layer may further enhance SOC (\ref{fig:SC-Conv}\textcolor{blue}{d}),
leading to more efficient conversion. Similarly, non-local voltages,
corresponding to enhanced SOC, have been observed on graphene decorated
with small amounts of adatoms or nanoparticles\citep{Balakrishnan2013},
and attributed to skew scattering on enhanced SOC perturbations\citep{Ferreira2014}.
Furthermore, it is appealing to explore other 2D systems that possess
larger intrinsic SOC, such as layered transition-metal dichalcogenides
(TMDs)\citep{Yankowitz2015}. Spin pumping and spin-transfer torque
experiments on TMDs\citep{Lee,Wang2015a}, notably MoS$_{2}$ and
WS$_{2}$, have shown promise for spin\textendash charge conversion.
Whereas semiconducting TMDs have limited utility for spintronics,
owing to their small mobility and large resistivity, heterostructures
of large-SOC TMDs and high-mobility graphene might be more useful.
Considering the large SOC enhancement of graphene in proximity with
WS$_{2}$\citep{Wang2015a}, larger spin\textendash charge conversion
effects can be expected in such structures. Heterostructures of the
TI Bi$_{2}$Te$_{2}$Se and graphene have also shown efficient transfer
of spin current to the graphene layer\citep{Vaklinova}.

%% file: SOCRev_C-Magnetism.tex
\section*{Interfacial Spin Interactions \protect \\
and Chiral Magnetism}

\noindent 
\begin{figure*}
\begin{centering}
\includegraphics[width=6.9in]{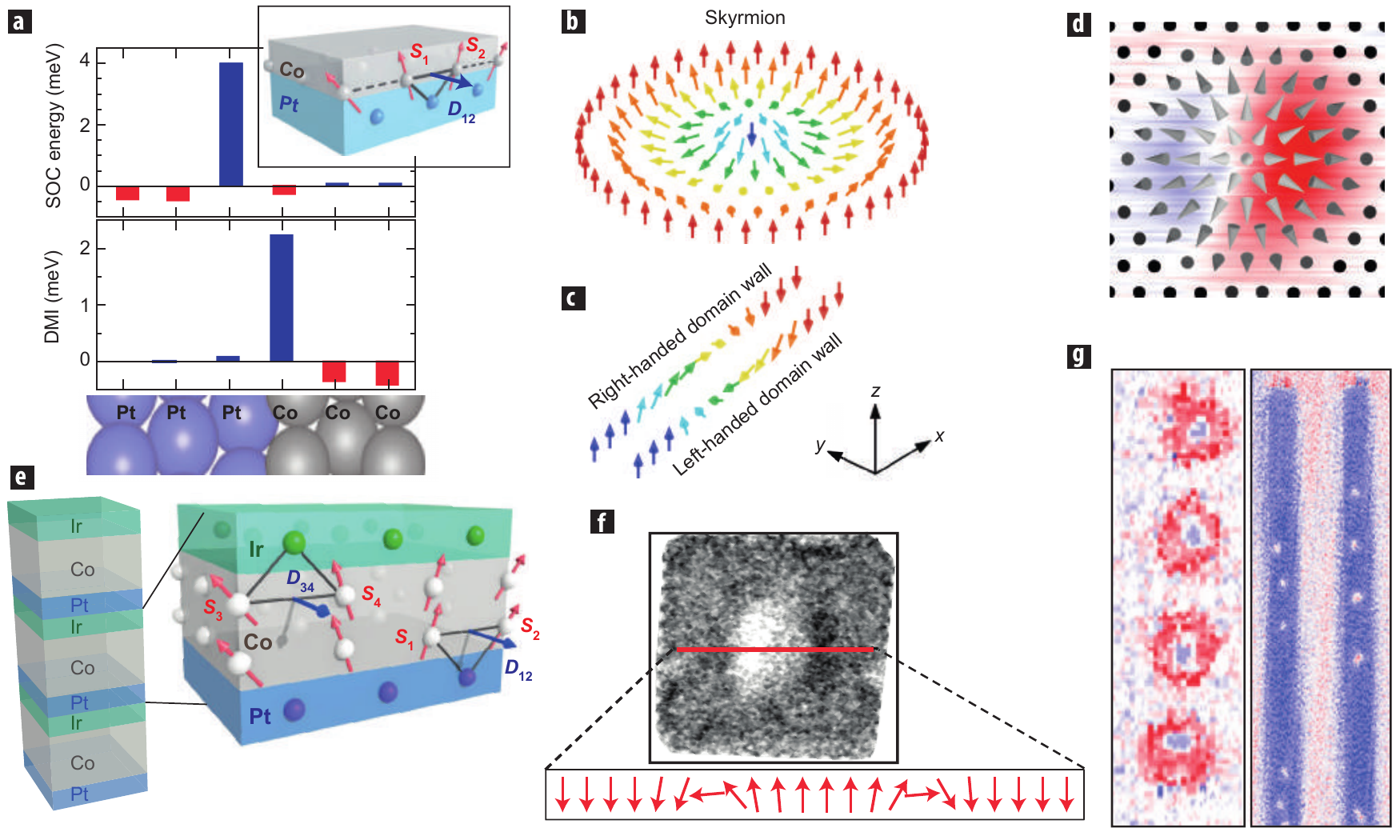}
\par\end{centering}
\caption{\textbf{Interfacial DMI and Chiral Spin Textures.} (a) Anatomy of
interfacial DMI from \emph{ab initio} calculations\citep{Yang2015a}.
Bottom, Layer-resolved DMI in a Pt/Co bilayer. Top, distribution of
SOC energies associated with the DMI in the interfacial Co layer.
Inset, a schematic of DMI at the interface between a ferromagnetic
metal with out-of-plane magnetization (Co, grey) and a strong SOC
metal (Pt, blue)\citep{Fert2013}. The DMI vector $\boldsymbol{D}_{12}$,
associated with the triangle composed of two Co atoms and a Pt atom,
is perpendicular to the plane of the triangle. $\boldsymbol{S}_{1,2}$,
neighbouring spins. \textbf{(b, c)} Schematics of the spin configuration
in interfacial-DMI-induced chiral spin textures such as magnetic skyrmions
(b) and chiral Néel DWs (c), with the colour scale corresponding to
the out-of-plane magnetization component. \textbf{(d)} SP-STM imaging
of an individual skyrmion (with a diameter of 8~nm at a field of
3.25~T) in a Fe/Pd bilayer on Ir(111)\citep{Romming2013}, acquired
in constant-current topographic mode, with an in-plane magnetized
tip, with the modelled magnetization overlaid (arrows). \textbf{(e)}
Skyrmion stabilization in multilayers, illustrated using a multilayer
stack of Ir/Co/Pt\citep{Moreau-Luchaire2015a}. The close-up of the
trilayer shows DMI vectors ($\boldsymbol{D}_{12}$ and $\boldsymbol{D}_{34}$)
at the top (Co/Ir) and bottom (Pt/Co) interfaces of Co. The effective
DMI magnitude is enhanced by the same direction of $\boldsymbol{D}_{12}$
and $\boldsymbol{D}_{34}$ at the different interfaces. \textbf{(f)}
RT skyrmions in a Pt/Co/MgO multilayer in a lithographed 400~nm \texttimes{}
400~nm square, seen by XMCD-PEEM\citep{Boulle2016}, with the magnetization
profile along the red line shown below. \textbf{(g)} RT skyrmions
in (Ir/Co/Pt) \texttimes 10 multilayers patterned into 300-nm-diameter
disks (left) or 200-nm-wide tracks (right), seen by STXM\citep{Moreau-Luchaire2015a}.
\label{fig:Sk-Form}}
\end{figure*}

\noindent \textsf{\textbf{Interfacial DMI}}

\noindent Magnetic materials that lack inversion symmetry can host
the DMI in the presence of strong SOC. Consequently, neighbouring
spins tilt with respect to each other, leading to spatial modulations
of the spin orientation. If the magnitude of $\boldsymbol{D}_{12}$
(\ref{eq:DMI}) is sufficiently large, then the competition between
the `winding\textquoteright{} DMI and `aligning\textquoteright{} exchange
interactions can give rise to non-collinear ground states\citep{Bode2007,Bogdanov1994,Bogdanov2001,Muhlbauer2009}.
Such chiral spin structures were initially identified in non-centrosymmetric
single crystals\citep{Bogdanov1994,Muhlbauer2009,Yu2010a,Nagaosa2013}.
However, of increasing scientific interest and technological relevance,
is their manifestation in films and multilayers with interfacial DMI
(\ref{fig:Sk-Form}\textcolor{blue}{a}).

Interfaces between ultrathin magnetic materials and metals with strong
SOC can host DMI, owing to broken inversion symmetry\citep{Fert1990}.
Large interfacial DMI (effective magnitude of $\boldsymbol{D}_{12}$
comparable to the exchange constant $J$) and ensuing spin textures
were first observed in ultrathin epitaxial magnetic films: spin spirals
in Mn on W(110)\citep{Bode2007}, and skyrmions on Fe and Fe/Pd on
Ir(111)\citep{Heinze2011,Romming2013}. In the presence of out-of-plane
anisotropy ($K$), this also leads to collinear magnetic domains separated
by Néel-type chiral DWs, that is, those with the chirality of spin
rotation through the DW determined by the orientation of $\boldsymbol{D}_{12}$
(\ref{fig:Sk-Form}\textcolor{blue}{c})\citep{Heide2008,Meckler2009a,Chen2013e,Emori2013}.
More recently, it has been possible to demonstrate the presence of
DMI-induced skyrmions at RT in magnetic multilayers grown by sputtering\citep{Moreau-Luchaire2015a,Woo2015,Jiang2015},
opening the possibility for using such structures in spintronics technologies\citep{Fert2013}.

The schematic in \ref{fig:Sk-Form}\textcolor{blue}{a} (inset) shows
such an interface between a magnetic film (Co) and a metal with strong
SOC (Pt). \emph{Ab initio} studies have provided insights into the
mechanism for interfacial DMI across several materials\citep{Heinze2011,Dupe2014,Yang2015a}.
For instance, it has been shown that, at a Co/Pt interface, the DMI
is strongest in the Co layer that is closest to the interface, and
relatively negligible in other Co layers (\ref{fig:Sk-Form}\textcolor{blue}{a},
bottom)\citep{Yang2015a}. Furthermore, the energy source of the large
DMI in the interface Co layer is located predominantly in the neighbouring
large-SOC Pt layer (\ref{fig:Sk-Form}\textcolor{blue}{a}, top), adding
credence to a direct correspondence between the DMI and interfacial
electronic states. The DMI in such multilayers has been measured directly
using spin-wave dispersion mapping techniques, and recent studies
have shown its inverse relationship with the thickness of the magnetic
layer \textendash{} a clear signature of its interfacial nature\citep{Cho2015a}.

\noindent \textsf{\textbf{Magnetic Skyrmions in Thin Films}}

\noindent The large DMI observed in these magnetic multilayers is
of particular interest because it induces new chiral spin textures,
known as magnetic skyrmions. Skyrmions are quasi-2D spin textures
wherein the out-of-plane magnetization is inverted at the centre and
rotates smoothly with a fixed chirality across its width (\ref{fig:Sk-Form}\textcolor{blue}{b}).
Skyrmions are distinguished by the topological number $S$ that characterizes
the winding of the normalized local magnetization $\boldsymbol{m}(\boldsymbol{r})$
(at position $\boldsymbol{r}$) in 2D systems:

\noindent 
\begin{equation}
S=\frac{1}{4\pi}\,\int d\boldsymbol{r}\,\boldsymbol{m}(\boldsymbol{r})\cdot\left(\partial_{x}\boldsymbol{m}(\boldsymbol{r})\times\partial_{y}\boldsymbol{m}(\boldsymbol{r})\right)=\pm1\label{eq:Sk-Winding}
\end{equation}

\noindent This topological number indicates that the skyrmion magnetization
covers the entirety ($4\pi$) of the unit sphere\citep{Nagaosa2013}.
Even though skyrmion-like objects (such as magnetic bubbles) can be
stabilized without DMI (for example, by dipolar interactions), they
would not have a fixed topological number $S$, which is crucial to
skyrmion properties\citep{Bogdanov2001,Nagaosa2013}. In fact, the
non-trivial topology of the skyrmion affords an energy barrier, protecting
its spin structure: the spin configuration cannot be twisted continuously
to obtain a different $S$ (for example, that of uniform polarization).
Another key property of the magnetic skyrmion is its solitonic nature
with a finite extension: it can move as a particle for as long as
it can be stabilized in a uniform ferromagnetic background.

\begin{figure*}
\begin{centering}
\includegraphics[width=4.6in]{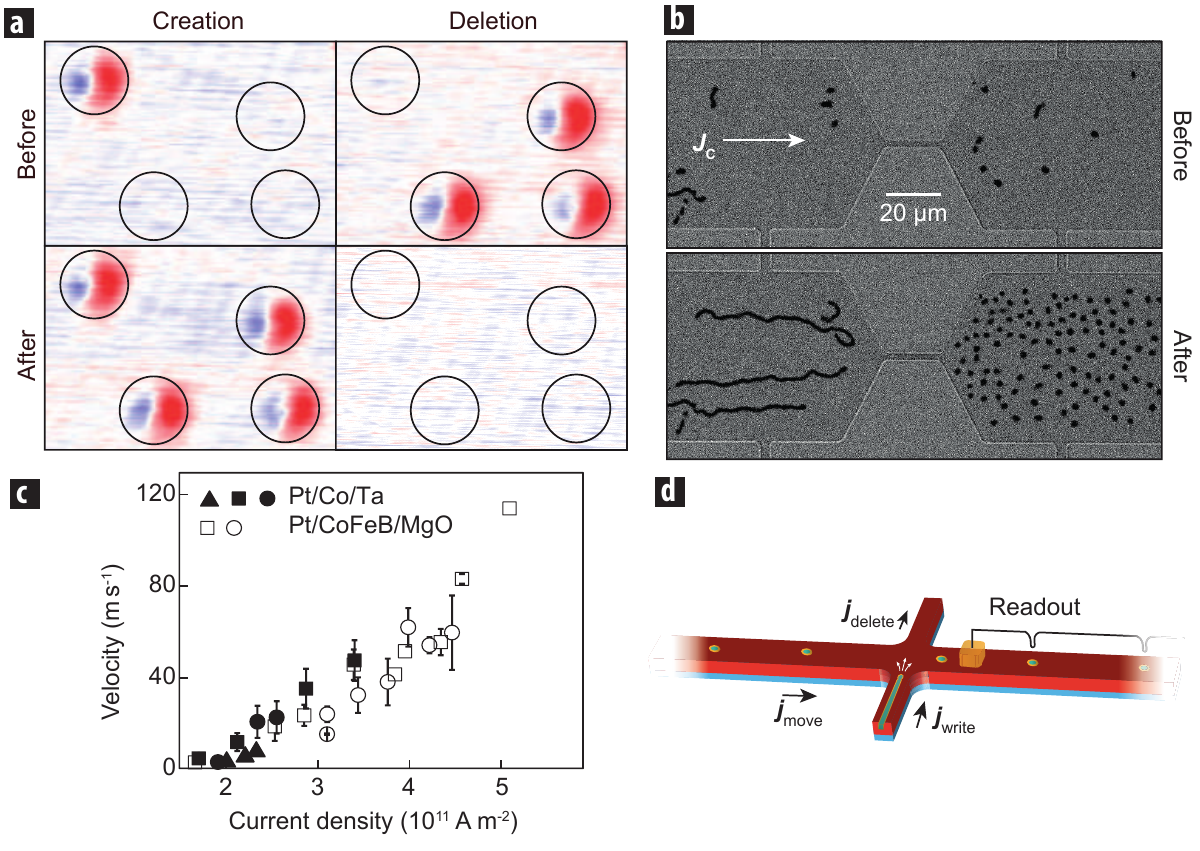}
\par\end{centering}
\caption{\textbf{Manipulation of Magnetic Skyrmions. (a)} Individual skyrmions
(with diameters of 8~nm at a field of 3.25~T) in Fe/Pd/Ir(111) before
and after SP-STM manipulation\citep{Romming2013}, demonstrating the
creation and annihilation of individual skyrmions at specific locations.
\textbf{(b)} Skyrmions in a Ta/CoFeB/TaO$_{x}$ structure\citep{Jiang2015},
before (top) and after (bottom) applying a current pulse through a
constriction, with current-induced nucleation and subsequent motion
of several skyrmions, as seen by magneto-optical Kerr effect (MOKE)
microscopy. \textbf{(c)} Experimental measurement of the current-induced
skyrmion velocity in tracks of Pt/Co/Ta and Pt/CoFeB/MgO\citep{Woo2015}
multilayers using STXM. \textbf{(d)} Schematic of a skyrmion-based
memory device\citep{Bergmann2015} in which skyrmions could be deleted,
moved and written by the corresponding current $\boldsymbol{j}$ .\label{fig:Sk-Manip}}
\end{figure*}

Skyrmions in ultrathin magnetic films were first observed in epitaxial
magnetic monolayers on heavy metal substrates (Fe on Ir(111)), for
which $D_{12}/J$ can be extremely large (about 1)\citep{Heinze2011,Romming2013}.
Here, skyrmions form a stable lattice configuration and the large
value of $D_{12}/J$ results in spin rotation over shorter length
scales, reducing the skyrmion size to just a few atomic lengths. However,
the formation of skyrmions herein requires low temperatures (about
10~K). Isolated skyrmions can be stabilized in a metastable state
by applying a magnetic field, for example in Fe/Pd on Ir(111) (\ref{fig:Sk-Form}\textcolor{blue}{d})\citep{Romming2013},
or at zero field if the DMI is sufficiently large, yet smaller than
the threshold value for stabilizing a skyrmion lattice or a spin spiral\citep{Rohart2013,Sampaio2013},
that is, if $D<4\sqrt{AK}$, where $D$ is the normalized DMI per
unit area of the film, $A$ is the exchange stiffness and $K$ characterizes
the out-of-plane anisotropy. Such individual skyrmions can further
be nucleated or annihilated \ref{fig:Sk-Manip}\textcolor{blue}{a}).
Owing to their topological protection, they are highly stable\citep{Hagemeister2015},
but only in applied field and at low temperature, which limits their
use as individual particles in devices. 

A crucial challenge for device applications is the RT stabilization
of such small individual skyrmions. The Néel or Curie temperatures
of bulk materials that are known to host skyrmions are generally below
RT. In addition, skyrmions in ultrathin magnetic monolayers (see,
for example, \ref{fig:Sk-Form}\textcolor{blue}{d}) are stable only
at low temperatures. A prominent direction towards practical RT skyrmions
is the development of multilayers with additive interactions at successive
interfaces\citep{Moreau-Luchaire2015a,Woo2015}. First, the interfacial
DMI can be enhanced by an appropriate choice of elements forming the
multilayer stack. For example, in the case of Ir/Co/Pt multilayers,
the sign of $D$ is opposite for Ir/Co and Pt/Co interfaces. As a
result, when Ir and Pt layers are on opposite sides of the Co layer,
their effects are additive, thus increasing the net DMI magnitude\citep{Moreau-Luchaire2015a}.
Furthermore, such trilayer stacks can be repeated (for example, the
ten repeats of (Ir/Co/ Pt); \ref{fig:Sk-Form}\textcolor{blue}{e}),
and skyrmions in successive stacks are coupled through ultrathin non-magnetic
layers, leading to columnar RT skyrmions ($\ge30$~nm) stabilized
by their larger magnetic volume\citep{Moreau-Luchaire2015a,Woo2015}.
RT skyrmions have also been found in magnetic bilayers (\ref{fig:Sk-Form}\textcolor{blue}{f},
\ref{fig:Sk-Manip}\textcolor{blue}{b}), although generally with a
larger diameter\citep{Jiang2015,Chen2015,Boulle2016,Gilbert2015}.
These efforts offer promising directions towards stack engineering
of magnetic interactions to tune skyrmion properties in films for
device applications\citep{Soumyanarayanan2016b}.

\noindent 

\noindent \textsf{\textbf{Detection and Manipulation of Chiral Spin
Textures}}

Skyrmions in epitaxial films were first imaged using spin-polarized
scanning tunnelling microscopy (SP-STM; \ref{fig:Sk-Form}\textcolor{blue}{d})\citep{Heinze2011,Romming2013}.
Since then, they have been imaged in sputtered multilayer films using
various magnetic microscopy techniques, including scanning transmission
X-ray microscopy (STXM; \ref{fig:Sk-Form}\textcolor{blue}{g})\citep{Moreau-Luchaire2015a,Woo2015},
photoemission electron microscopy (PEEM; \ref{fig:Sk-Form}\textcolor{blue}{f})\citep{Boulle2016},
spin-polarized low-energy electron microscopy (SPLEEM)\citep{Chen2015},
and magneto-optical Kerr effect (MOKE) microscopy (\ref{fig:Sk-Manip}\textcolor{blue}{b})\citep{Jiang2015}.
Importantly, skyrmions can also be detected using a variety of thermodynamic
and transport techniques\citep{Neubauer2009}. In particular, the
Berry phase that is accumulated by electrons traversing the 2D spin
texture of skyrmions results in an additional component in anomalous
Hall effect measurements, known as the topological Hall effect\citep{Nagaosa2013,Neubauer2009}.
The Hall signal can be used to detect the presence of skyrmions and
to address their motion in films and devices\citep{Neubauer2009,Schulz2012}.
However, such Hall signatures of skyrmions have been detected thus
far only in bulk crystal and films with intrinsic DMI\citep{Neubauer2009,Schulz2012,Huang2012a};
these techniques remain to be established in multilayer films with
interfacial DMI.

Magnetic skyrmions, owing to their small size and non-trivial topology,
are attractive candidates for data storage in magnetic materials \textendash{}
provided that they can be nucleated, moved and read. Several nucleation
techniques have been explored with micromagnetics simulations\citep{Sampaio2013,Iwasaki2013b}.
In SP-STM experiments on Fe/Pd bilayers (\ref{fig:Sk-Manip}\textcolor{blue}{a}),
individual skyrmions were nucleated and deleted using the current
injected from the STM tip\citep{Romming2013}. In other experiments,
skyrmions have been created by applying field pulses\citep{Woo2015}.
A remarkable result in this regard is the recent demonstration of
``blowing of skyrmion bubbles''\citep{Jiang2015,Heinonen2016}, generated
by the current divergence out of a constriction (\ref{fig:Sk-Manip}\textcolor{blue}{b}).
In future, skyrmions should be able to be moved with notable ease
compared with, for example, DWs\citep{Schulz2012} by exploiting the
SOT provided by the spin current\citep{Sampaio2013,Tomasello2014,Kang2016},
which emerges naturally from the spin Hall effect of the neighbouring
heavy metal layers. The dynamic properties of skyrmions have been
explored using micromagnetics simulations and microscopy techniques
in device configurations\citep{Woo2015,Jiang2015}. These works demonstrate
that skyrmions can be manipulated with current and field pulses in
lithographed geometric structures (\ref{fig:Sk-Manip}\textcolor{blue}{b,
c})\citep{Woo2015,Jiang2015} \textendash{} techniques that can be
incorporated in memory devices with relative facility.

These properties of magnetic skyrmions portend great potential towards
realizing high-density and energy-efficient memory\citep{Tomasello2014,Kang2016}.
Several applications and architectures have been proposed and modelled,
including skyrmion-based memory devices (\ref{fig:Sk-Manip}\textcolor{blue}{d})
analogous to DW based racetrack memory\citep{Parkin2008}. The interest
in skyrmions with respect to DWs is the smaller current that is needed
for their displacement and the weaker influence of defects on skyrmion
motion\citep{Sampaio2013}. More efficient SOT could also be obtained
by using the spin\textendash charge conversion at the interfaces of
TIs instead of the spin Hall effect. 

DMI of intermediate strength has direct relevance to chiral DWs\citep{Heide2008},
which are also being actively explored\citep{Chen2013e,Emori2013,Yang2015}.
The motion of magnetic DWs under SOT depends on the relative configuration
of the DW magnetization and the type of SOT under consideration\citep{Khvalkovskiy2013}.
The large DW velocity observed in perpendicular anisotropy films deposited
on heavy metals was initially understood as emerging from the SOT
at the interface between the magnet and the heavy metal\citep{Moore2008}.
However, it has been recently understood that the marked enhancement
in DW velocity results from the stabilization of chiral, Néel-type
DWs by interfacial DMI\citep{Thiaville2012}. This DMI-induced stabilization
suppresses the Walker breakdown mechanism that typically limits DW
dynamics and explains the efficient SOT action on this type of DW\citep{Emori2013}.
In addition, the DW chirality, which is determined by DMI, corresponds
to a fixed direction of motion; hence, all chiral DWs move in the
same direction in a given stack structure. 

The motion of DMI-stabilized chiral DWs can lead to new realizations
of nanoscale data storage. For example, consider a memory element
that stores the information by using the DW position. The DW can have
two stable positions (for example, using a notch along a short stripe).
Here, the magnetization switch can be measured using a magnetic tunnel
junction\citep{Fukami2009}. Such memory architectures based on switching
nanoscale spin structures require much less current than does conventional
MRAM, wherein the magnetization of the entire device needs to be reversed.
Furthermore, the fast SOT-induced motion of chiral DWs\citep{Yang2015}
is relevant to advancing the development of other concepts, such as
DW racetrack memory\citep{Parkin2008}.

%% file: SOCRev_D-Impact.tex
\section*{Applications and Outlook}

The interplay between SOC and inversion-symmetry breaking has given
rise to fascinating phenomena at surfaces and interfaces, especially
in the past decade. TIs have been described as a ``new state of quantum
matter''\citep{Hsieh2008}, and the emergence of interfacial DMI has
given rise to non-trivial spin structures. However, the remarkable
properties induced by SOC at surfaces and interfaces go beyond chiral
magnets and spin topology, including several types of 2D materials
with intrinsic or engineered SOC. The emergent characteristics of
these SOC-induced phenomena, which are robust at RT, offer several
potential applications.

First, spin\textendash momentum locking in TIs can be exploited, via
their interaction with normal metals, to obtain unprecedented efficiency
in spin\textendash charge conversions. High spin\textendash charge
conversion efficiency will probably be harnessed in future spintronic
devices, such as SOT-MRAM or nano-batteries. Second, the protection
from backscattering in TIs can be used in low-dissipation devices.
The topologically protected skyrmion spin configurations ultimately
represent the smallest achievable size for an emergent non-volatile
magnetic memory element in magnetic films, with immediate relevance
to information storage. Skyrmions can be moved, created and annihilated
in nanostructures, making them suitable for `abacus\textquoteright -type
applications such as racetrack memory. 

Although several of these avenues for application have emerged only
in the past decade, the rapid advances along this front make us optimistic
about the time frame in which we can reasonably expect to see devices
that realize the potential of SOC-induced properties. Consider skyrmions
in magnetic multilayers: pioneering efforts on epitaxial films at
low temperature\citep{Heinze2011,Romming2013} were soon followed
by RT observations in the kind of sputtered multilayer films\citep{Moreau-Luchaire2015a,Woo2015,Jiang2015,Boulle2016}
typically used in spintronics technologies. Demonstrations of their
small size, electrical nucleation and motion \textendash{} all under
ambient conditions \textendash{} offer further technological promise.
Other devices include skyrmion-based transistors\citep{Zhang2015},
oscillators and microwave detectors\citep{Finocchio2005}. Furthermore,
concepts of SOT-MRAM could be extended to utilize skyrmions.

The applications that could emerge from TIs may evolve over a range
of time frames. Fundamental materials challenges remain to be overcome,
including the integration of TI compounds into existing elements of
spintronic technologies and the control of defects that impede their
exotic properties at RT. However, recently discovered TIs demonstrate
unprecedented spin\textendash charge conversion efficiency at RT\citep{Rojas-Sanchez2016}.
They are therefore suitable candidates to replace spin-Hall-effect
heavy metals \textendash{} for example, for memory state switching
of SOT-RAM \textendash{} in the near future\citep{Mellnik2014}. Another
exciting prospect is the use of TIs to generate efficient SOT for
manipulating skyrmions and chiral DWs. Charge-to-spin conversion by
Bi/Ag Rashba interfaces has been proposed as a way to develop non-volatile
and complementary metal-oxide-semiconductor (CMOS)-compatible spin
logic devices\citep{Manipatruni2015}, and spin-to-charge conversion
in ferromagnet\textendash TI devices can be used for microwave-driven
spin batteries\citep{Mahfouzi2010} and electrical power generators
that exploit spin currents induced by temperature gradients\citep{Cahaya2014}.
Furthermore, the high fidelity of RT spin-polarized photocurrents
generated on TIs\citep{McIver2011} offer promising opto-spintronics
applications, including transparent conducting electrodes, phase modulators
and polarizers. Finally, upon overcoming the materials challenges,
we foresee that the ambipolarity of topological surface states could
be used to create low-power spintronics devices such as a topological
$p\lyxmathsym{\textendash}n$ junction\citep{Wang2012i}. One may
envisage a device where an applied magnetic field turns such a $p\lyxmathsym{\textendash}n$
junction into an electronic Mach\textendash Zender interferometer\citep{Ilan2015},
enabling a tunable junction transmission with spin filtering properties. 

The discovery of the novel states induced by SOC and inversion-symmetry
breaking at surfaces and interfaces opens up such a broad perspective
that the introduction of their topological properties will have a
definitive and substantial effect on the technology of spintronics.